\documentclass[12pt,preprint]{aastex}
\usepackage{epsfig}
\usepackage{psfig}

\newcommand{\chisq}{$\chi^2$}

\newcommand{\CXO}{{\sl CXO}}

\newcommand{\EG}{3EG J2020+4017}
\newcommand{\RX}{RX J2020.2+4026}
\newcommand{\CG}{2CG078+2}

\shorttitle{{A multi-wavelength search for a counterpart of the
unidentified  $\gamma$-ray source 3EG J2020+4017 (2CG078+2)}}
\shortauthors{W.~Becker et al.}

\begin{document}

\title{{A multi-wavelength search for a counterpart of the
unidentified  $\gamma$-ray source 3EG J2020+4017 (2CG078+2)}}

\author{
Werner Becker\altaffilmark{1},
Martin C. Weisskopf\altaffilmark{2},
Zaven Arzoumanian\altaffilmark{3},
Duncan Lorimer\altaffilmark{4},\\
Fernando Camilo\altaffilmark{5},
Ronald F. Elsner\altaffilmark{2},
Gottfried Kanbach\altaffilmark{1},
Olaf Reimer\altaffilmark{6},\\
Douglas A. Swartz\altaffilmark{7},
Allyn F. Tennant\altaffilmark{2},
Stephen L. O'Dell\altaffilmark{2}
}

\altaffiltext{1}
{Max Planck Institut f\"ur Extraterrestrische Physik, 85741 Garching bei M\"unchen, Germany}
\altaffiltext{2}
{Space Sciences Department, NASA Marshall Space Flight Center, SD50, Huntsville, AL 35812}
\altaffiltext{3}
{USRA, Laboratory for High-Energy Astrophysics, Goddard Space Flight Center, Greenbelt, MD 20771}
\altaffiltext{4}
{University of Manchester, Jodrell Bank Observatory, Macclesfield, Cheshire SK11 9DL, UK}
\altaffiltext{5}
{Columbia Astrophysics Laboratory, Columbia University, 550 West 120th Street, New York, NY 10027}
\altaffiltext{6}
{Ruhr-Universi\"at Bochum, 44780 Bochum, Germany}
\altaffiltext{7}
{USRA, Space Sciences Department, NASA Marshall Space Flight Center, SD50, Huntsville, AL 35812}

\begin{abstract}
In search of the counterpart to the brightest unidentified gamma-ray source 3EG J2020+4017
(2CG078+2) we report on new X-ray and radio observations of the $\gamma$-Cygni field with the
{\em Chandra} X-ray Observatory and with the Green Bank Telescope (GBT). We also report on 
reanalysis of archival ROSAT data. With {\em Chandra} it became possible for the first 
time to  measure the position of the putative gamma-ray counterpart \RX\ with sub-arcsec 
accuracy and to deduce its X-ray spectral characteristics. These observations demonstrate 
that RX J2020.2+4026 is associated with a K field star and therefore is unlikely to be the 
counterpart of the bright gamma-ray source 2CG078+2 in the SNR G78.2+2.1 as had been previously 
suggested. The {\em Chandra} observation detected 37 additional X-ray sources which were 
correlated with catalogs of optical and infrared data. 
Subsequent GBT radio observations covered the complete 99\% EGRET likelihood contour of 
3EG J2020+4017 with a sensitivity limit of $L_{820} \approx 0.1\, \mbox{mJy kpc}^2$ 
which is lower than most of the recent deep radio search limits. 
If there is a pulsar operating in \EG, this sensitivity limit suggests that the 
pulsar either does not produce significant amounts of radio emission or that its
geometry is such that the radio beam does not intersect with the line of sight.
Finally, reanalysis of archival ROSAT data leads to a flux upper limit of
$f_x(\mbox{0.1-2.4 keV}) < 1.8 \times 10^{-13}\, \mbox{erg s}^{-1}\,\mbox{cm}^{-2}$
for a putative point-like X-ray source located within the 68\% confidence contour 
of \EG. Adopting the SNR age of 5400 yrs and assuming a spin-down to X-ray energy 
conversion factor of $10^{-4}$ this upper limit constrains the parameters of a 
putative neutron star as a counterpart for \EG\, to be $P$ {\small $\gtrsim$} 
$160\, (d/1.5\, \mbox{kpc})^{-1}\mbox{ms}$, $\dot{P}$ {\small $\gtrsim$} $5\times 
10^{-13}\, (d/1.5 \mbox{kpc})^{-1}\,\mbox{s s}^{-1}$ and $B_\perp$ {\small $\gtrsim$} 
$9\times 10^{12}\, (d/1.5\,\mbox{kpc})^{-1}$ G.

\end{abstract}


\section{Introduction}

The error boxes of unidentified gamma-ray sources are usually large, and thus
the task of finding appropriate candidate counterparts at other wavelengths 
has not been easy. About 20 bright point-like gamma-ray sources were found near 
the Galactic plane using COS-B (Swanenburg et al.~1981) some of which may be 
concentrations of molecular hydrogen (Mayer-Hasselwander \& Simpson 1990). 
Another few, such as the Crab  and Vela pulsars, were identified based on 
their periodic emission (Thompson et al. 1975).  The nature of the other 
objects remained unknown.

The much more sensitive EGRET telescope on the {\em Compton Gamma-Ray Observatory}
({\em CGRO}) was expected to contribute decisively to the identification 
of the COS-B sources.  And indeed, the higher count rates and tighter source 
locations provided by EGRET confirmed the existence of most of the COS-B 
sources and led to the identification of several other sources. Most prominent 
were the gamma-ray pulsars Geminga and PSR~B1706--44 which could be identified 
on the basis of detections at X-ray and radio wavelengths (see e.g.~Kanbach 2002 
and Becker \& Pavlov 2001 for a review and references).  At high Galactic latitudes  
about 90 new high-energy sources could be correlated with blazars. The final EGRET 
catalog of gamma-ray sources lists 271 objects (Hartman et al.~1999) of which about 
170 are unidentified. Their distribution suggests that most of them are Galactic. 
The origin and nature of this population of extremely energetic objects is
clearly of interest.

Seven of the Galactic gamma-ray sources are rotation-powered  pulsars,
identified through the periodic modulation of their gamma-ray fluxes. 
These seven are also persistent, point-like sources at gamma-ray energies. 
In the 100 MeV -- 1 GeV energy range, these sources have hard, power-law-like
spectra with high-energy cut-offs at a few GeV. Although rotation-powered 
pulsars are best known as radio sources, this is not true for all --- Geminga, 
for example, shows at best marginal evidence of pulsed radio emission (Kuzmin 
\& Losovkii 1997). Geminga is thus taken as the prototype of a `radio-quiet' 
gamma-ray pulsar of which many more should exist in the Galaxy. 
Although Geminga's gamma-ray luminosity is rather low (its small distance of
about 160 pc makes it a bright source) the property of radio faintness could be
indicative of pulsar emission where the beamed radiation at different
wavelengths is emitted into widely different directions. Such a model may be 
applicable to young, high luminosity pulsars as well (Yadigaroglu \& Romani 1995).
There are other models to explain Geminga's radio faintness though (e.g.~Gil et al.~1998).

A review of the spectra of unidentified low Galactic latitude EGRET
sources (Bertsch et al. 2000; Merck et al. 1996) shows that about 10
objects exhibit the very hard power-law type spectra with a cut-off at
several GeV as seen also in the identified pulsars. These objects would 
be prime targets for identification efforts at other wavelengths. 
Relatively deep radio searches (at 770 MHz) at the positions of several of these 
sources have not found radio counterparts (Nice \& Sayer 1997). 
Population studies of the unidentified gamma-ray sources close to the Galactic
plane indicate that their luminosities are also quite compatible with the
luminosities of the younger identified pulsars (Kanbach et al.~1996). 
Suggestions, other than pulsars, for the nature of these gamma-ray 
sources have also been widely discussed. Energetic objects, like massive young 
stars or OB associations and SNRs have been correlated with the 3EG catalog 
(e.g. Romero et al. 2000) and certainly indicate a close relationship with 
the gamma-ray sources. 

Multi-wavelength observations focusing on promising candidate sources have been
quite successful in recent years. Observations in X-rays have been useful: 
e.g.~in the cases of 3EG~J2006$-$2321 = PMN2005$-$2310 (Wallace et al.~2002) and 
3EG J2016+3657 = B2013+370 (Halpern et al.~2001a). New pulsar/isolated 
neutron star identifications were reported, e.g.~3EG~J2227+6122 (Halpern et 
al.~2001b) by discovery of the characteristic pulsar period of RX/AX~J2229.0+6114.
X-ray observations were used to relate the high Galactic latitude source
3EG~J1835+5918 to an isolated neutron star, RX~J1836.2+5925 (Reimer et al.
2001; Mirabal \& Halpern 2001; Halpern et al. 2002). 

With the wealth of incoming discoveries from the Parkes multi-beam pulsar survey, 
promising associations between newly discovered radio pulsars and EGRET sources 
have also been discussed. These associations include the two young pulsars 
PSR~J1420--6048 and PSR~J1837--0604 (D'Amico et al.~2001) in the vicinity of 
3EG~J1420--6038 and 3EG~J1837--0606, respectively; PSR~J1016--5857 near 
the SNR G284.3--1.8 is a plausible counterpart for 3EG~J1013--5915 (Camilo et 
al.~2001).  In a recent survey of 56 unidentified EGRET 
sources Roberts et al.~(2004) found a radio pulsar located inside the 95\% 
likelihood map in six of the investigated gamma-ray sources. The discovery
of PSR J2021+3651 in the error box of GeV 2020+3658 using the 305m Arecibo
radio telescope is another positive example (Roberts et al.~2002; 2004; Hessels 
et al.~2004). However, Torres, Butt \& Camilo (2001), and more recently Kramer 
et al.~(2003), who have summarized the observational status of the radio 
pulsars and EGRET-detected gamma-ray sources concluded that, in many cases, 
further multi-frequency investigations are required in order to conclusively 
translate a proposed association into a final source identification.

\EG\ is among the brightest persistent sources in the EGRET sky. Originally 
listed as a COS-B source (2CG$078+01$) it is still unidentified. Its gamma-ray 
flux is consistent with constant flux (Hartman et al.~1999), and the spectrum 
is hard and best described by a power-law with photon-index of $1.9\pm 0.1$. 
Merck et al.~(1996) found evidence for a spectral break at $\sim 4$ GeV which 
has been confirmed in recent studies by Bertsch et al.~(2000) and Reimer \& Bertsch (2001).

Examining all archival EGRET data and using photons $>1$ GeV, Brazier et
al.~(1996) found a best position at $\alpha_{2000} = 20^{\rm h} 20^{\rm m} 
15^{\rm s}$, $\delta_{2000} = 40^\circ 21'$ with a $20\times 14\,
\mbox{arcmin}^2$ 95\%-confidence error box.  This position was consistent 
with the 2EG catalog position and placed the EGRET source within the 
$\gamma$-Cygni supernova remnant G78.2+2.1. 

The remnant G78.2+2.1 consists of a $1^\circ$-diameter, circular radio shell
with two  bright, broad opposing arcs on its rim (Higgs, Landecker \& Roger
1977; Wendker, Higgs \& Landecker 1991). G78.2+2.1 has a kinematic distance of 
$1.5$ kpc $\pm\,30\%$ (Landecker, Roger \&  Higgs 1980; Green 1989) and is 
estimated to have an age of 5400 yr (Sturner \& Dermer 1995). A very bright star, 
$\gamma$-Cygni ($m_v=2.2$, spectral type F8Iab) lies on the eastern edge and 
lends its name to the remnant. A small H{\sc ii} region, located 
close to the star, forms the so-called $\gamma$-Cygni nebula.

\begin{figure}
\begin{center}
\epsfig{file=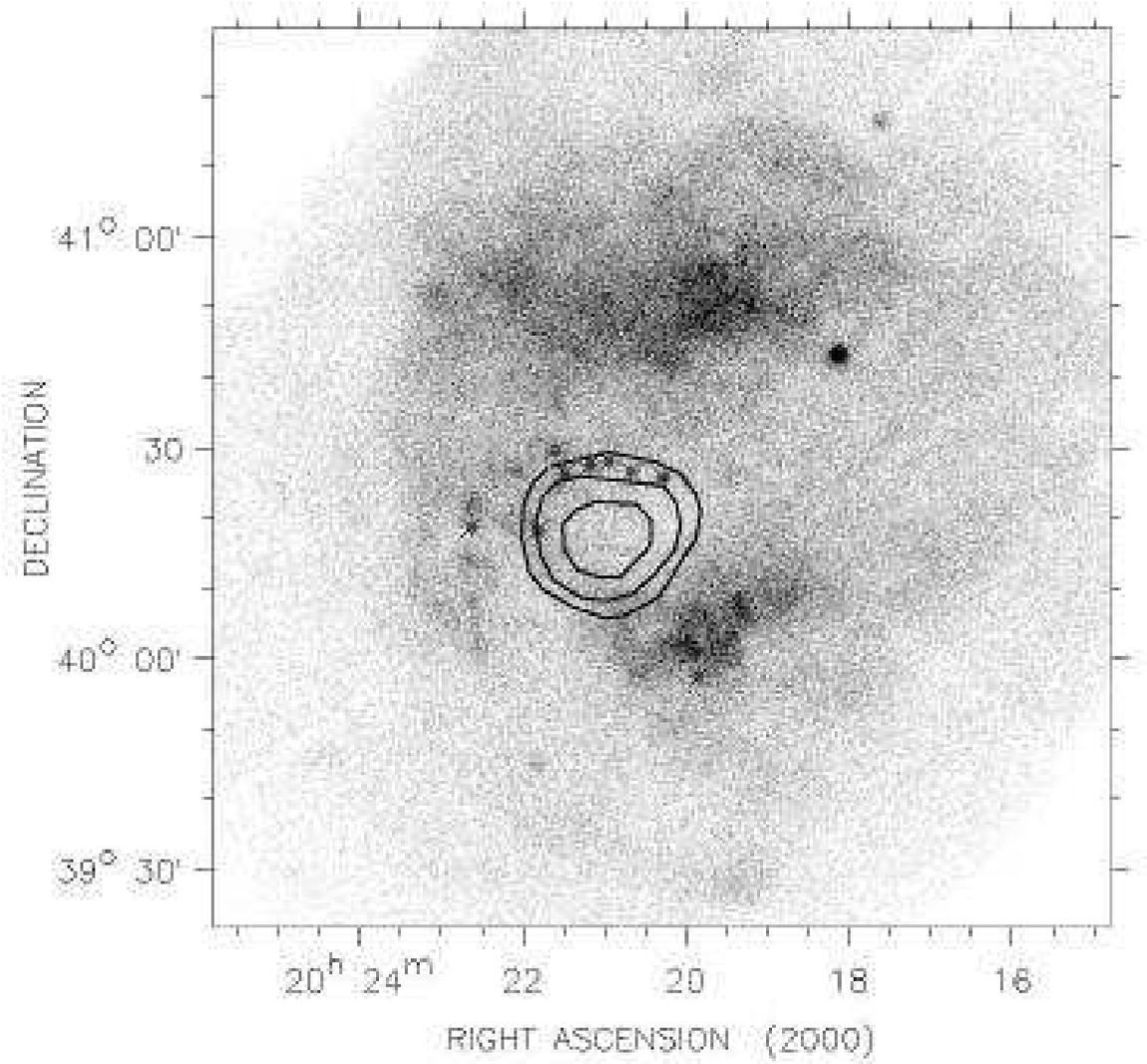,width=16cm,angle=0}
\end{center}
\figcaption{
 ROSAT PSPC field around \EG.  The $68\%$, $95\%$, and $99\%$ contour lines from the 
3EG EGRET likelihood map and the ROSAT HRI sources detected in our re-analysis of archival
ROSAT data (see Table \ref{hri_sources}) are indicated. As in the radio band, the SNR is 
dominated by two bright arcs on the northern and southern edges.
\label{f:rosat}}
\end{figure}

Brazier et al.~(1996) analyzed ROSAT PSPC data viewing the $\gamma$-Cygni region. 
Six PSPC observations were targeted at celestial positions within 40 arcmin of the
EGRET source.  
These ROSAT observations are combined and shown in Figure~\ref{f:rosat}.
The point source RX J2020.2+4026 is located within the 95\% confidence 
contour of the 2EG position of \CG\, (Brazier et al.~1996), and was 
suggested by these authors to be the X-ray counterpart to the gamma-ray source.
Brazier et al.~(1996) and Carraminana et al.~(2000) provided a possible optical
counterpart for RX J2020.2+4026.  Optical follow-up observations revealed a 
14.5 magnitude K0V star nearby and within the $\approx 6''$ ROSAT error 
circle. The X-ray to optical flux ratio of this star was found to be marginally 
consistent with that found for late-type stars (Stocke et al.~1991; Fleming et 
al.~1995), so that an association of RX J2020.2+4026 with the gamma-ray source 
could not be excluded (Brazier et al.~1996).

With the 3EG catalog (Hartman et al.~1999), an improved  position of
\CG\, became available: $\alpha_{2000} = 20^{\rm h} 21^{\rm m} 1\fs0$,
$\delta_{2000} = 40^\circ 17' 48''$,  i.e.~shifted in right ascension 
and declination by a few arc-minutes with respect to the 2EG position 
used by Brazier et al.~(1996).  With this improved position the  proposed 
counterpart RX J2020.2+4026 is no longer located within the 95\%  contour 
of \EG. The 99\% likelihood contour, however, still includes 
\RX\, (Figure~\ref{f:rosat}).

In this paper we report on follow-up studies of \RX\ with {\em Chandra} and
the Green Bank Radio Telescope. The {\em Chandra} observations were taken with 
the aim to determine the position and spectrum of \RX\ with high precision 
and to explore the possible connection with \EG. GBT observations at 820 MHz 
were made in order to search the EGRET error box of \EG\ for a young radio
pulsar.

\section{{\em Chandra} Observations and Data Analysis \label{obs}}

Our 30 ksec {\em Chandra} observation (ObsID 3856) was taken on 2003 January 26
using three Advanced CCD Imaging Spectrometer (ACIS) CCDs (S2,3,4) in the faint, 
timed-exposure mode with a frame time of 3.141 s.  
Standard {\em Chandra} X-ray Center (CXC) processing (v.6.8.0) has applied
aspect corrections and compensated for spacecraft dither. 
Level~2 event lists were used in our analyses. 
Events in pulse invariant channels corresponding to 0.5 to 8.0 keV were
selected for the purpose of finding sources on S2 and S4 (after destreaking). 
The energy range 0.25 to 8.0 keV was utilized for source finding with S3.  
Due to uncertainties in the low energy response, data in the range 0.5 to 8.0
keV were used for spectral analysis.  There were no instances of increased background. 

\begin{figure}
\begin{center}
\epsfig{file=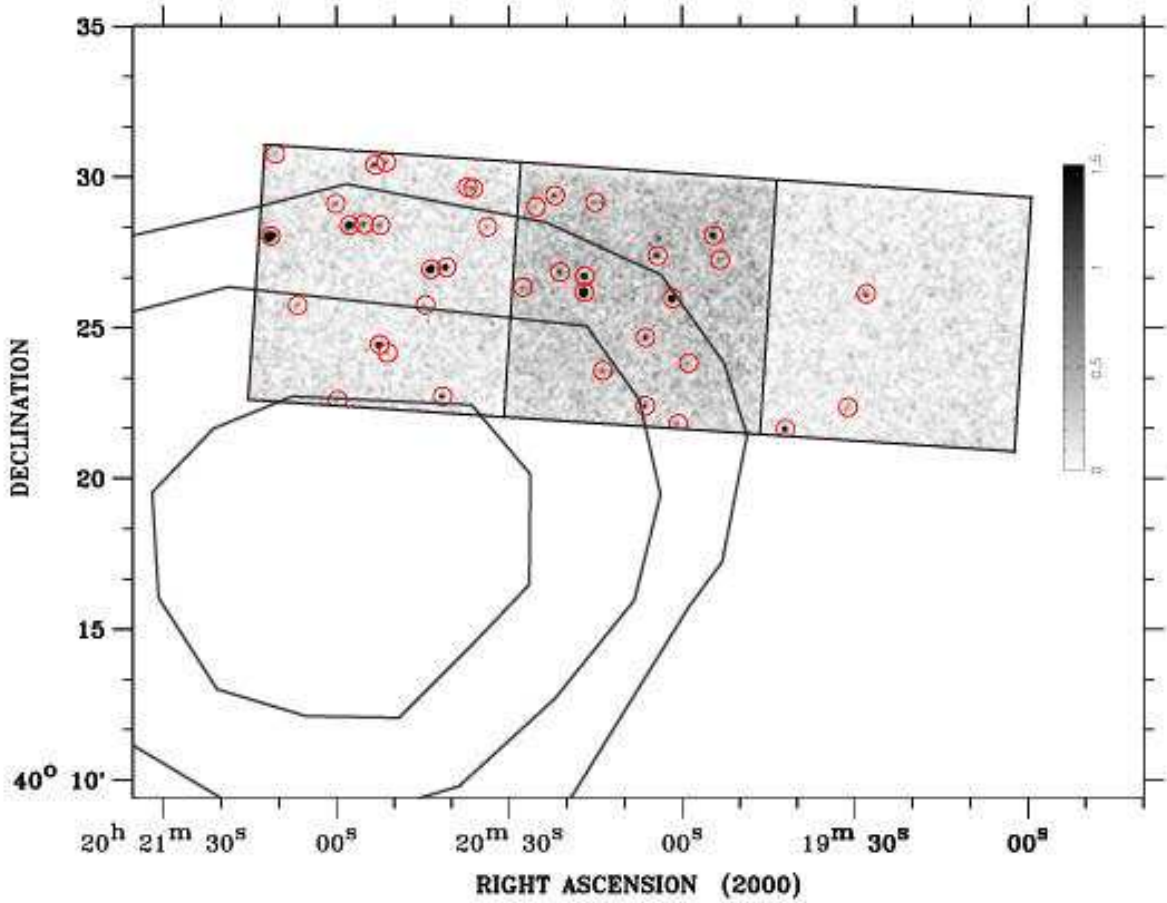,width=16cm,angle=0}
\end{center}
\figcaption{
{\em Chandra} ACIS field (chips S2, 3, 4 from left to right, square panels) of \EG. 
The $68\%$, 95\% and $99\%$ contour lines from the 3EG EGRET likelihood-map are
shown as well. The positions of the 38 {\em Chandra} sources listed in Table
\ref{GamCyg_source_table} are indicated.
\label{f:chandra}}
\end{figure}

The center position used for the {\em Chandra} pointing was that of the ROSAT source
RXJ2020.2+4026, at $\alpha_{2000} = 20^{\rm h}\,20^{\rm m}\,17\fs0$ and
$\delta_{2000} = 40\degr\, 26\arcmin\, 9''$.  The ACIS image overlaid 
with the 3EG likelihood contour lines of \EG\, is shown in Figure \ref{f:chandra}. 
The positions of \RX\, and 37 other X-ray sources detected by {\em Chandra} are 
also indicated. \RX\, still appears to be the brightest X-ray source in the field.
 
\subsection{Image Analysis \label{image_analysis}}

We used the same source finding techniques as described in Swartz et al.~(2003)
with the circular-Gaussian approximation to the point spread function, and a
minimum signal-to-noise ratio (S/N) of 2.6, expected to result in much less 
than 1 accidental detection in the field. The corresponding background-subtracted 
point source detection limit is $\sim$10 counts. Nineteen sources were found 
on the S2 chip, 16 on S3, and three on the S4 CCD. 

Table~\ref{GamCyg_source_table} lists the 38 X-ray sources. 
To simplify the discussion and to show the CCD in which a source was detected
(the S2,4 and S3 chips have different sensitivities) sources are denoted as
SXYY with X=2,3,4 indicating the CCD and YY=01,02,03\ldots indicating the
ordering in RA.

Table~\ref{GamCyg_source_table} gives the source positions, the associated uncertainty in these
positions, and the signal-to-noise ratio.
The positional uncertainty listed in column~7
is given by $r=1\farcs51(\sigma^2/N + \sigma^2_o)^{1/2}$ where $\sigma$ is the size
of the circular Gaussian that approximately matches the point-spread function (PSF) 
at the source location, $N$ is the vignetting-corrected number of source counts, 
and $\sigma_o$ represents the systematic error. The factor 1.51 sets the radius 
to that which encloses 68\% of the circular Gaussian. The Table also lists 
potential counterparts in either the United States Naval Observatory Catalog 
(USNO-B1.0; Monet et al. 2003), the Two Micron All Sky Survey (2MASS), or 
the various ROSAT catalogs.  With the exception of the target source, ROSAT 
sources were listed if their position was within $5\farcs0$ of the {\em Chandra} 
position. Uncertainties in the plate scale\footnote{See
http://asc.harvard.edu/cal/Hrma/optaxis/platescale/} imply a systematic
uncertainty of $0\farcs13$, and, given the differences in the
USNO and 2MASS and \CXO\ positions, we have used $0\farcs3$ as a 
conservative estimate for $\sigma_o$.

The non-X-ray candidate counterparts were selected by searching a circular region 
centered on the X-ray position and whose radius was the 99\%-confidence radius 
(3.03/1.51 times the positional uncertainty listed in column 7) continuing the 
assumption that the point-spread function is described by a circular Gaussian. 
We do recognize, of course, that the assumption is not accurate far off-axis, 
however, this is partially compensated in that there is a good deal of 
conservatism built into the definition of the positional uncertainty and by the 
fact that there is very little impact on the position centroid. 

As a further check on our estimate of the systematic error in the correlation
of the X-ray and optical positions a position-error weighted least squares
fit was performed. Right ascension, declination, and roll angle of the
pointing position were taken as free parameters of this fit. The uncertainties
in the optical positions were taken from the USNO-B1.0 catalog. The uncertainty
in the X-ray position was that discussed previously, including the systematic
uncertainty $\sigma_0 = 0\farcs3$. There are 26 optical candidates selecting
only one of the two possible counterparts to S401. The fit was excellent
($\chi^2$ of 55 for 52 degrees of freedom), independent of which candidate was
associated with S401. The fitted quantities for right ascension, declination,
and roll angle were $-0\farcs12\pm0\farcs12, 0\farcs01\pm0\farcs10$, and
$-32\arcsec\pm65\arcsec$, respectively. Considering that we have ignored any
possible systematic errors in the non-X-ray positions, such as those due to
proper motion, we feel that applying this offset is unjustified.
 
\subsection{Optical and Infrared Counterparts\label{o_ir_counterparts}}

The BROWSE\footnote{See
http://heasarc.gsfc.nasa.gov/db-perl/W3Browse/w3browse.pl.} 
feature was used to search for cataloged objects at or near the X-ray 
positions listed in Table~\ref{GamCyg_source_table}. 
All available BROWSE catalogs were selected to be interrogated and the 
99\%-confidence region around the X-ray sources were searched for possible 
counterparts. 

\subsubsection{USNO-B1.0\label{USNO}}


There are 3387 USNO-B1.0 sources in a $12'$ radius centered on the X-ray pointing
direction. To the extent that these are uniformly distributed there are
$2.1 \times 10^{-3}\,\mbox{sources arcsec}^{-2}$ and this density was used to 
calculate the expected average number of accidental coincidences listed in column
4 of Table~\ref{t:counterparts}.  The probability of getting one or more matches 
by chance is given by the Poisson probability $1 - \exp(-N_{r99})$ which for small
values is approximately $N_{r99}$. These probabilities are always below 7\% and 
most (20 of 26) below 2\%.  The separation between the X-ray source and the 
optical source is listed in the 8th column of Table~\ref{GamCyg_source_table}.
The position of the candidate optical counterpart is also listed in  
Table~\ref{t:counterparts}. There are two optical candidate counterparts for S401. 

The position of the source S312 is found to match that of the optical K0V-star
which has a USNO-B1.0 position of $\alpha_{2000}=20^{\rm h}\,20^{\rm m}\,17\fs13$ 
and $\delta_{2000}=+40\degr\, 26\arcmin\, 14\arcsec9$.

\subsubsection{2MASS\label{2MASS}}

There are 5061 2MASS sources in the $12'$ radius circle centered on the pointing
position and the inferred density was used to calculate the probability of an 
accidental coincidence that is listed in column 7 of Table~\ref{t:counterparts}. 
The probabilities are always below 12\% and about half of them are below 2\%. 
Other pertinent information concerning the potential infrared counterparts is listed 
in Tables~\ref{GamCyg_source_table} and \ref{t:counterparts}. 
In Table~\ref{t:counterparts} we have also listed, where relevant, the separation 
between the optical and the infrared candidate counterparts. 
In all cases but S209, these separations are sub-arcsecond, implying, 
apart from the exception, that the optical and the infrared sources are the same. 

\begin{figure}
\begin{center}
\epsfig{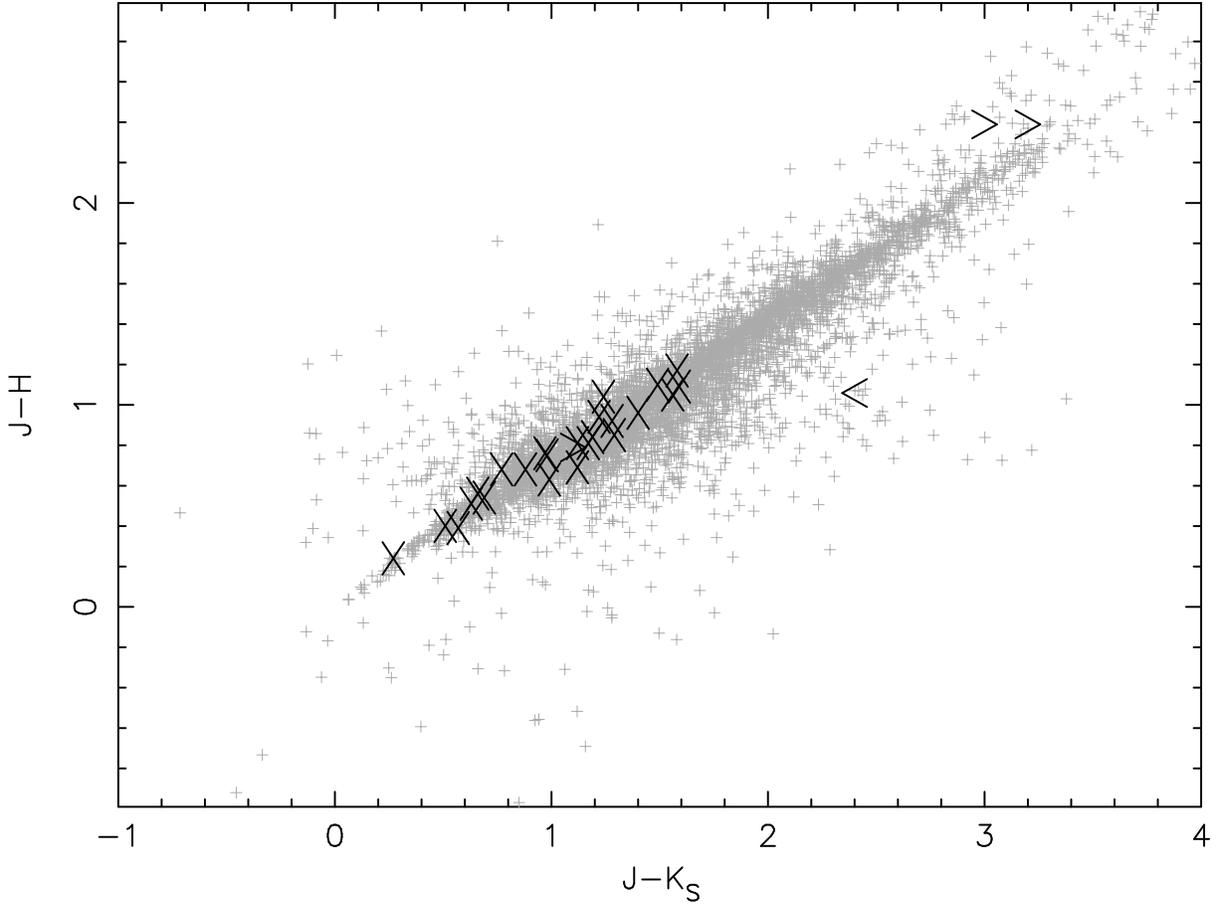}
\end{center}
\figcaption{Color-color diagram for 2MASS field stars located within a $12'$
radius from the center of the field of view. The colors of 2MASS objects 
associated with X-ray sources are indicated with either an ``X", or a ``$>$" 
in the case where there are only upper limits (see Table~\ref{t:2MASS_colors}).
\label{f:2MASS_colors}}
\end{figure}

Table~\ref{t:2MASS_colors} shows the magnitudes and colors of the 2MASS counterparts.
Figure~\ref{f:2MASS_colors} shows the colors of all of the 2MASS sources in a
12-arcmin radius circle centered on the pointing direction. 
With two possible exceptions, the inferred counterparts of the X-ray sources 
appear to be distributed as the field sources. Those that do not appear to 
be reddened have colors of moderately late-type stars. This is not surprising 
since the $\gamma$-Cygni field is close to the Galactic plane where approximately 
90\% of 2MASS sources are stars\footnote{
http://www.ipac.caltech.edu/2mass/releases/second/doc/}. Although most 
Galactic-plane 2MASS objects are normal stars, the majority of objects 
identified with X-ray sources need not be stars. For example, the X-ray 
emission may arise from a compact companion. The distribution of the colors 
of the X-ray selected subset, however, seems to reproduce the distribution of 
the field objects. The two most reddened sources are S208 and S309 which have 
J$-$K$_S > 3$ and thus may possibly be background sources (AGNs) absorbed by 
the Galactic column.

The position of the source S312 is found to match that of the optical K0V-star
which has a 2MASS position of $\alpha_{2000}=20^{\rm h}\,20^{\rm m}\,17\fs13$ 
and $\delta_{2000}=+40\degr\, 26\arcmin\, 14\farcs5$.

\subsection{Spectral Analysis\label{spectral_analysis}}

Point-source counts and spectra were extracted from within the radii listed in
column 4 of Table~\ref{GamCyg_source_table}.  The background estimation was
determined from creating data sets for each CCD after removing the events 
from each source  region out to a radius 10 times the extraction radius 
listed in Table~\ref{GamCyg_source_table}.

Only a few of the 38 detected sources have sufficient counts to warrant an
attempt at an individual spectral analysis. 
In descending order of the number of detected counts, these are sources S312
(253 cts), S206 (213 cts), S219 (201 cts), S305 (176 cts), S214 (101 cts), and
S204 (85 cts).  All spectral analyses used CIAO 3.0.2 to 
extract the pulse invariant (PI) files and CXC CALDB 2.25 calibration files 
(gain maps, quantum efficiency uniformity and effective area) to generate the 
appropriate effective area and response functions. The spectral data were 
corrected for the effects of charge transfer inefficiency produced by 
proton damage to the front-illuminated CCDs early in the mission. 
Finally, we accounted for the impacts of molecular contamination
on the ACIS filters with the number of days since launch set at 1275. 
For the absorbing column we used TBABS in XSPEC (v.11.2) with the default 
cross sections but with the abundances set to Wilms, Allen \& McCray (2000).
The data were binned with no less than 10 counts per 
spectral bin. All given errors on spectral parameters are extremes on the two 
interesting parameters at the 68\% confidence range. As noted previously, 
spectral analysis were restricted to the energy range $0.5-8.0$ keV because 
of the large uncertainties in the ACIS spectral response at low energies. 

\subsubsection{Source S312 (\RX) \label{S312}}

Three spectral models were applied. The first model, an absorbed power law, 
resulted in a statistically excellent fit ($\chi^2$ of 16.7 for 15 degrees of freedom) 
but with physically unreasonable parameters such as a very steep power law 
spectral index of 8.2.
 
\begin{figure}
\begin{center}
\epsfig{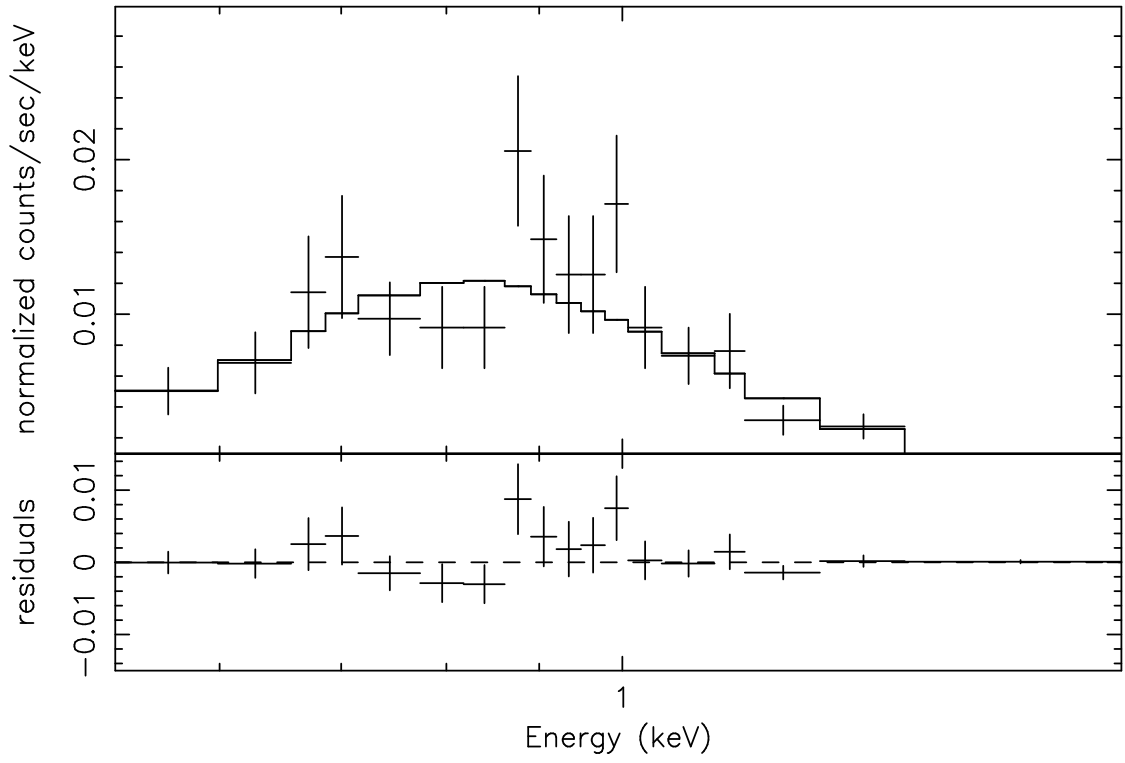}
\end{center}
\figcaption{Energy spectrum of source S312 fitted to an absorbed blackbody model
with correction for molecular contamination of the ACIS filter.
\label{f:spectrum_S312_bbody}}
\end{figure}

The second model was the thermal, emission-line XSPEC model {\tt mekal}.
With $Z=1.0\,Z_\odot$ (solar metallicity), we obtained an acceptable fit 
($\chi^2= 22.3$ for 15 degrees of freedom) with $N_{H}\approx 0.0\,\mbox{cm}^{-2}$, 
and $kT \approx 0.77$ keV.
The 1$\sigma$ confidence ranges for the column absorption and temperature are
$N_{H} = < 0.05 \times 10^{22}\,\mbox{cm}^{-2}$ and $kT= 0.68-0.83$ keV, 
respectively. The third model, a blackbody with $N_{H}\approx 0.4\times 
10^{22}\,\mbox{cm}^{-2}$  and $kT \approx 0.14$ keV, provides an alternative 
statistically acceptable representation of the X-ray  spectrum  (Table~\ref{t:spectra}). 
The spectrum and residuals of that fit are shown in Figure~\ref{f:spectrum_S312_bbody}.

From the H{\sc i} in the Galaxy (Dickey \& Lockman 1990) we compute the column absorption
through the Galaxy in the direction to $\gamma$-Cygni to be $N_{H} \sim 1.4\times 10^{22}\,
\mbox{cm}^{-2}$. This is comparable with the column absorption of sources in $\gamma$-Cygni
(Maeda et al.~1999; Uchiyama et al.~2002) and significantly higher than what is found from 
spectral fits of  S312, suggesting that this source is a foreground object. 

The low-column {\tt mekal} spectrum together with the positional identification would 
appear to establish S312 (\RX) as the X-ray counterpart to the K0V-star, invalidating
its association with the unidentified EGRET source \EG. This conclusion is bolstered by 
considering the upper limit to the X-ray luminosity of $< 3\times 10^{28}\,\mbox{erg s}^{-1}$ 
which we obtained from the X-ray flux of $f_x \sim 2.5 \times 10^{-14}\,\mbox{erg s}^{-1}
\mbox{cm}^{-2}$ and the distance upper limit of $< 356$ pc. The latter was derived from 
the star's spectral type K0 (Brazier et al.~1996) and distance modulus. We note that 
the luminosity is slightly high suggesting the star is rotating ``rapidly'' which is 
mildly inconsistent with the optical spectra.

\subsubsection{The other ``bright" sources \label{other_bright_sources}}

As with S312 (\RX) we binned the data to assure at least 10 counts per spectral 
bin and fit the data for the other relatively bright sources to the absorbed 
power law, {\tt mekal} and blackbody models. The results are summarized in Table~\ref{t:spectra}. 

\paragraph{S206 \label{S206}}
About 213 source counts were detected from this source.  The data clearly favor 
the power law spectrum with a photon-index in the 68\% confidence range
$1.67-2.36$. 

\paragraph{S219 \label{S219}}
This source is detected with 201 source counts near to the edge of the S2 CCD
and was seen by ROSAT (1RXH~J202111+402807). In contrast to S206, a more
complicated spectrum is called for.  We tried to fit the data with a
two-temperature {\tt mekal} model (not unusual for stars) which did  provide a
good fit ($\chi^2 = 15.0$ on 14 degrees of freedom).  The two best-fit
temperatures were 0.2 and 24 keV with the highly uncertain higher
temperature component providing only 3\% of the total flux. The best-fit
absorbing column was $N_{H}=0.7\times 10^{22}\,\mbox{cm}^{-2}$.

\paragraph{S305 \label{S305}} 

Similar to S312, all three spectral models provided statistically adequate fits
to the spectral data. 
In this case, however, it is not valid to argue that the power law index is
unphysical. 

\paragraph{S214 \& S204 \label{S214}} 

None of the three model spectral fits are acceptable (as with S219), 
however, with only a total of seven and five bins of spectral data respectively, 
it would not be surprising to be able to fit these data with a more 
complicated model. Both sources have both USNO-B1.0 and 2MASS counterparts. 

For the remaining 32 sources, all with fewer than 69 detected source 
counts, no spectral fitting was attempted.  These sources were, however, 
included in the X-ray color-color diagrams presented in the following section.

\subsubsection{X-ray color-color relation\label{ss:xray_color_color}}

\begin{figure}
\begin{center}
\epsfig{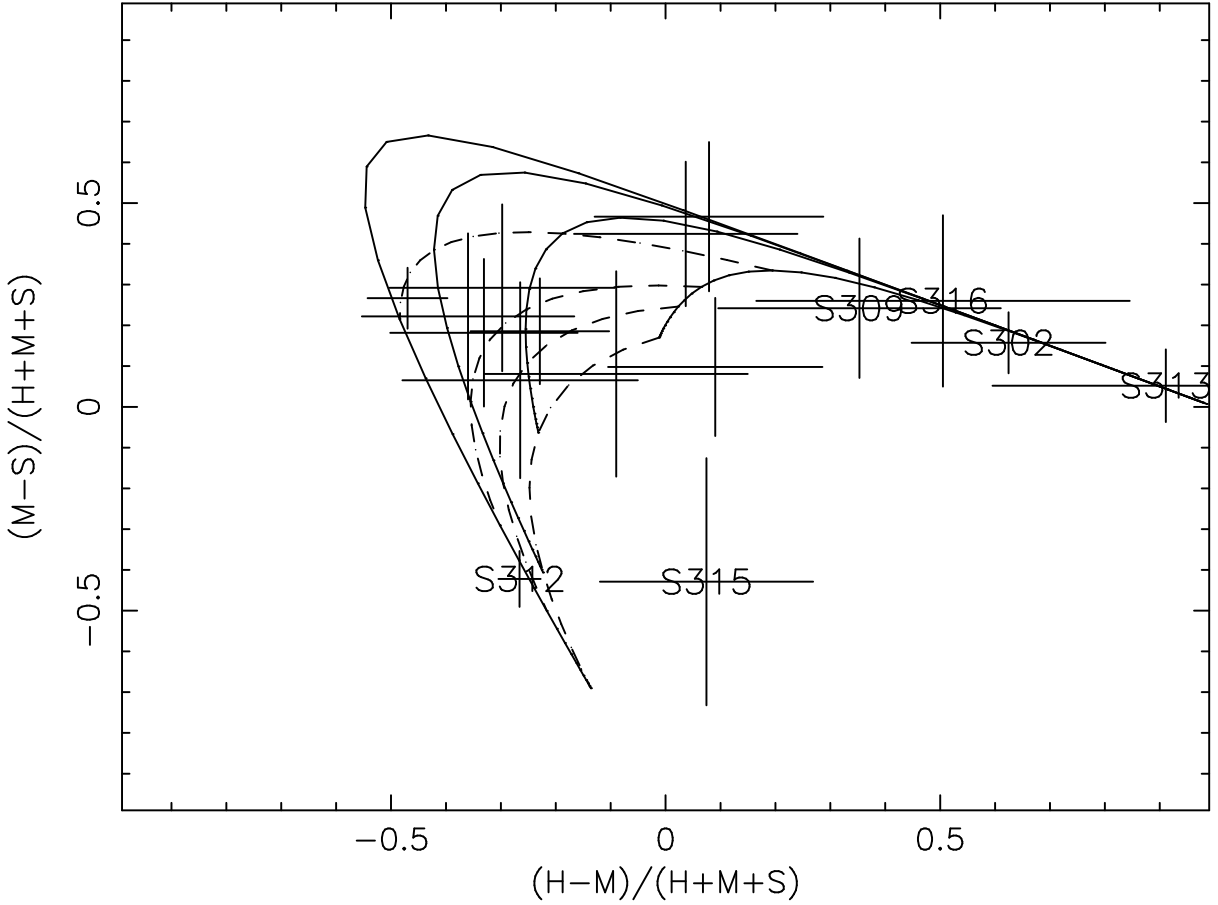}
\end{center}
\figcaption{X-ray ``color-color" diagram for the sources detected with the
back-illuminated CCD, S3. 
The bands are: S (0.5--1.0 keV); M (1.0--2.0 keV); H (2.0--8.0 keV). 
The solid lines are contours for power law spectra of constant photon number index ranging from $-1$ (innermost) to $-4$ (outermost) where $N_H$ is varying.
The dashed lines are contours of constant $N_H$ for a power law spectrum of varying spectral index. $N_H$ is 0.1, 1, 2, and 5 $\times 10^{21}$\,cm$^{-2}$ from the innermost to the outermost contour. 
Thus a source with spectral index $-1$ and $N_H$ of $10^{20}$\,cm$^{-2}$ would be placed on the plot at the intersection of the dashed and solid lines at approximately (0.0,0.15).
\label{f:S3bi_colors}}
\end{figure}

\begin{figure}
\begin{center}
\epsfig{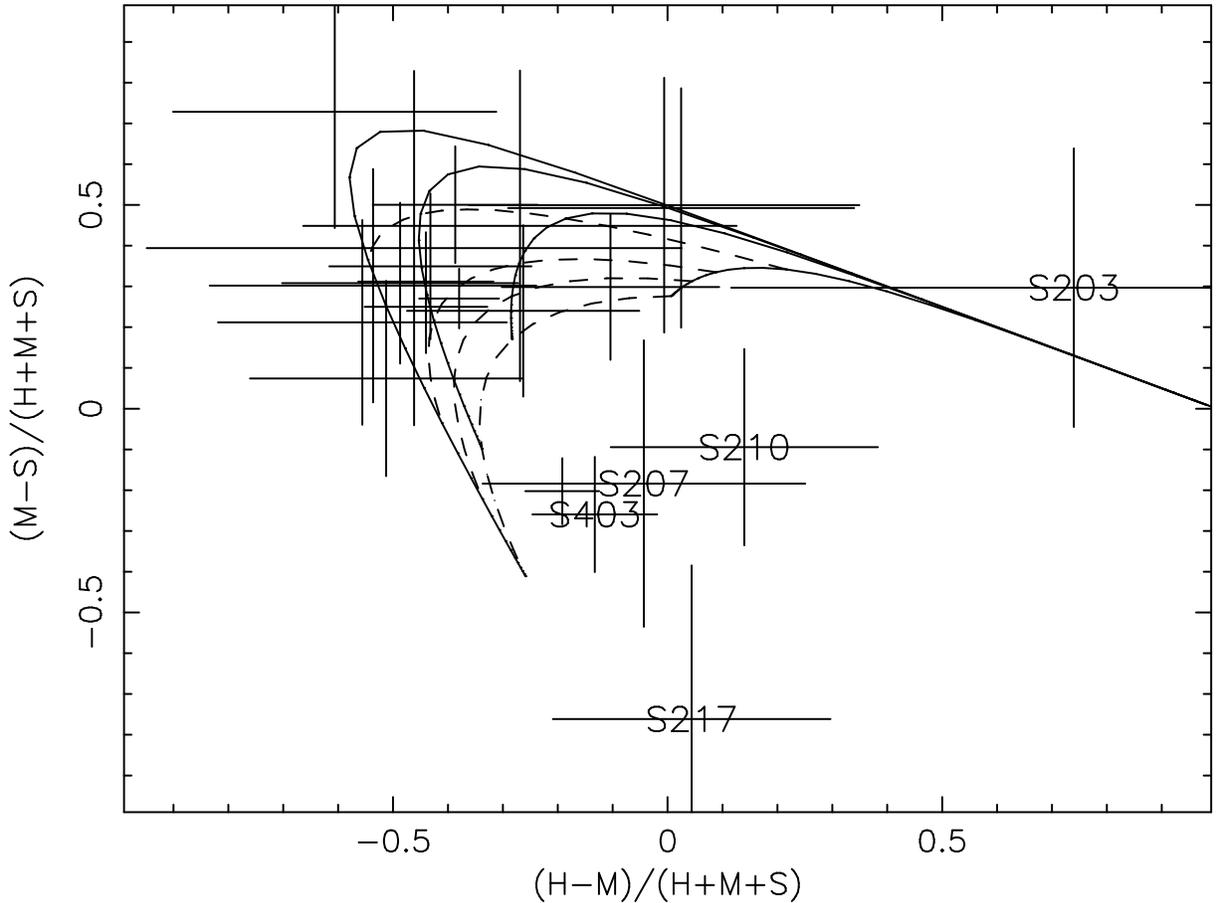}
\end{center}
\figcaption{X-ray ``color-color" diagram for the sources detected with the
front-illuminated CCDs, S2 and S4. 
The unlabeled point closest to S403 is from S219. 
The bands are: S (0.5--1.0 keV); M (1.0--2.0 keV); H (2.0--8.0 keV). 
The solid lines are contours for power law spectra of constant photon number index ranging from $-1$ (innermost) to $-4$ (outermost) where $N_H$ is varying.
The dashed lines are contours of constant $N_H$ for a power law spectrum of varying spectral index. $N_H$ is 0.1, 1, 2, and 5 $\times 10^{21}$\,cm$^{-2}$ from the innermost to the outermost contour. 
\label{f:S2fi_colors}}
\end{figure}

We show two X-ray ``color-color" diagrams in Figures~\ref{f:S3bi_colors} and
\ref{f:S2fi_colors}. Because of differences in the spectral responses, the
data were separated between back- and front-illuminated  CCDs, the 
back-illuminated CCDs being somewhat more sensitive to low energy X-rays. 
Clearly, since most of these sources were detected with a small number of
total counts, the uncertainties are large and it is difficult to draw any firm
conclusions. Thus the figures are primarily included for completeness. 
We note that the hardest sources (S203, S302, S313, S316), those that
occupy the upper-right portions of the diagrams, are among those with 
no USNO and/or 2MASS counterparts and these may be background sources 
(AGN) absorbed by the Galactic column. The very soft and unabsorbed sources 
(S315, S219, S403, S207, S210, S217) are likely to be associated with foreground 
stars and all have optical and infrared candidate counterparts. The 
infrared colors (Table~\ref{t:2MASS_colors}) of all of these objects, 
except for the candidate counterpart to S219, are those expected for 
evolved main sequence (primarily K) stars. The counterpart to S219 
would have to be a giant or supergiant. 

\subsection{Time Variability}

\begin{figure}
\begin{center}
\epsfig{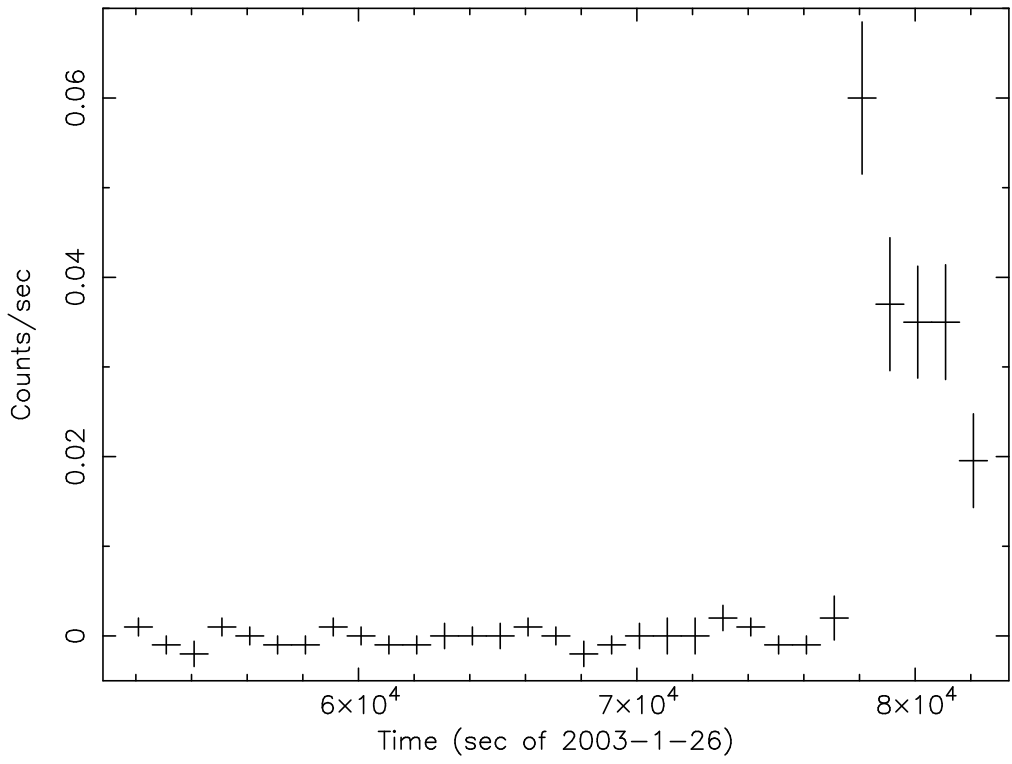}
\end{center}
\figcaption{Relative counting rate versus time for the source S206.
\label{f:s206flare}}
\end{figure}

The paucity of detected counts for the X-ray sources limit the ability to
draw many conclusions from the time series. In one case however, S206, the 
source was quiescent for most of the observation and then suddenly 
flared as shown in Figure~\ref{f:s206flare}.  There is also, less 
compelling, evidence that S219 flared. The flaring nature of S206
and a 2MASS counterpart are all consistent with coronal emission 
from a star. 

\subsection{Re-analysis of Archival ROSAT Data}

We have re-analyzed the archival ROSAT PSPC data used by Brazier et 
al.~(1996). Source detection algorithms (box car as well as maximum likelihood) 
found many sources, most of which appear to be associated with the diffuse 
emission of the remnant G78.2+2.1 as neither of these methods is ideally 
suited for searching for point sources embedded in extended sources 
(i.e.~patchy  background). The X-ray source \RX\, seen by Brazier et al.~(1996)
was the only point source detected. No other point source was found within the
\EG\, error box. Using these data we can thus set a $2\sigma$
count rate upper limit of $6.9 \times 10^{-4}$ PSPC counts/s for a putative
X-ray point source located within the region defined by the 68\% confidence 
contour of \EG. Assuming a power law spectrum with a photon 
index of two and a column absorption of $1.4\times 10^{22}\,\mbox{cm}^{-2}$
(Dicky \& Lockman 1990), the count rate upper limit corresponds to an energy
flux upper limit of $f_x(\mbox{0.1-2.4 keV}) < 1.8 \times 10^{-13}\,
\mbox{erg s}^{-1}\,\mbox{cm}^{-2}$ and $f_x(\mbox{0.5-8.0 keV}) < 1.7 \times 
10^{-13}\,\mbox{erg s}^{-1}\,\mbox{cm}^{-2}$, respectively. For a distance of 
1.5 kpc this yields $L_x(\mbox{0.1-2.4 keV}) < 4.8 \times 10^{31}\,
\mbox{erg s}^{-1}$ for the upper limit to the isotropic X-ray 
luminosity. 

In addition to the ROSAT PSPC observations, there are four HRI data sets 
in the ROSAT archive which partly cover the \EG\, error box. These data 
were taken between 1994 and 1997 with varying exposure times from 10 ksec 
to 36.5 ksec, respectively, and were not considered in the analysis 
of Brazier et al.~(1996). Table \ref{hri} list the observational details. 
A maximum-likelihood source-detection algorithm 
with a threshold of $5\sigma$ found six point sources in the data set 400899h, 
two in 202534h, one in 500339h and none in 202033h. The source properties 
are given in  Table \ref{hri_sources} and the positions are shown in Figure~\ref{f:rosat}.       

The region observed in the HRI observation 400899h mostly overlaps our 
{\em Chandra} observations. Of the six sources detected in this HRI observation, 
two new X-ray sources not seen by {\em Chandra} are detected; RX J202137.6+402959 
and RX J202057.8+402829. Thus, these sources appear to be variable. The other 
four HRI sources detected are S214, S219, 
S204 and S312 (the last is the putative X-ray counterpart of \EG\, 
proposed by Brazier et al.~1996). Among the other three newly detected 
sources, only RX J202150.5+401837 is located within the 95\% likelihood 
region for the position of 3EG J2020+4017. The other two sources are located far 
outside the 99\% contour and cannot account for an X-ray counterpart of 3EG J2020+4017.

\section{Radio Observations and Data Analysis \label{s:radio}}

\begin{figure}
\begin{center}
\epsfig{file=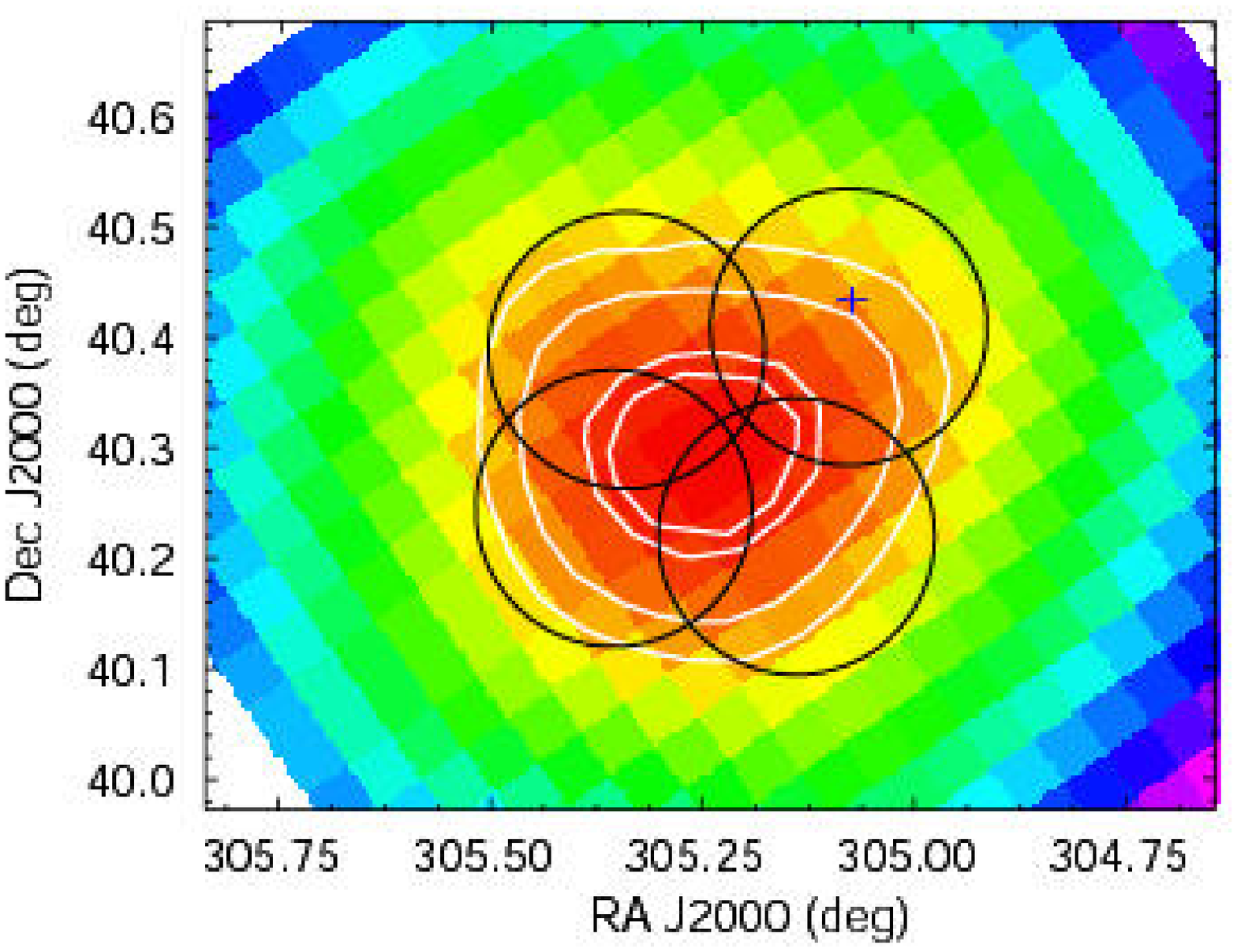,width=15cm}
\end{center}
\figcaption{Likelihood map ($> 100$ MeV) of \EG\ (see Hartman et al.~1999),
with smoothed contours superposed. The fields covered by the four radio observations are 
indicated by black circles. The cross indicates the position of \RX. \label{radio_pointings}}
\end{figure}

A deep search for radio pulsations from \CG\, was carried out 
using the GBT on 2003 December 27. The observations were made 
at a center frequency of 820 MHz using an identical data acquisition 
and analysis scheme as that described by
Camilo et al.~(2002) in their detection of 65-ms radio pulsations
from the pulsar in SNR~3C~58. Given the uncertain position of the
putative pulsar, the observing time was divided into four separate
pointings of the 15$'$ (FWHM) beam: one pointing was approximately centered
on RX J2020.3+4026, and the remainder were arranged so as to cover
much of the EGRET error region likely to contain the $\gamma$-ray
source with 99\% probability (Figure~\ref{radio_pointings}). Due 
to time constraints, dwell times of 4 hr for three of the pointings, 
and 3 hr for the fourth, were used.
The pointings are summarized in Table~\ref{radio_pointings_tabel}.  
Data were acquired with the Berkeley Caltech
Pulsar Machine (BCPM), an analog/digital filter bank (Backer et
al.~1997) that divides the frequency band into 96 contiguous channels
and samples the incoming voltages of the two orthogonal circular
polarizations received by the telescope every 72 $\mu$s. For these
observations, the channel bandwidth was set to 0.5 MHz so that a
total band of 48 MHz was recorded.  After summing the polarizations,
the resulting total-power time samples were written to disk with
4-bit precision for subsequent off-line processing. Known pulsars
were successfully detected before and after the \CG\, observations.

The data analysis was carried out at Jodrell Bank using standard
Fourier-based pulsar search software routines (for full details, see
Lorimer et al.~2000) which are freely
available\footnote{http://www.jb.man.ac.uk/$\sim$drl/seek}. In order to
reduce the volume of data, 12 (10) adjacent time samples were
added together prior to dedispersion of the 4-hr (3-hr)
observations. The resulting decimated time series had effective sampling times
of 864 and 720 $\mu$s.  This choice of decimation was convenient for the 
periodicity search, which uses a base-two Fourier transform algorithm.  For a
$2^{24}$-point Fourier transform, the effective integration times
were 3.9 hr and 3.2 hr. The latter data were zero padded.  Each
observation was analyzed separately. The data were first dedispersed
at 301 trial dispersion measure (DM) values between 0 and 300
cm$^{-3}$ pc. The expected DM from the NE2001 electron density model
(Cordes \& Lazio 2002) for $l=78.2\arcdeg$ and $b=2.1\arcdeg$ is $\approx 17$
cm$^{-3}$ pc, assuming a distance of 1.5 kpc (Landecker, Roger \&  Higgs 
1980). The maximum DM in this direction is 350 cm$^{-3}$ pc (Cordes \& 
Lazio 2002).
The resulting time series were then Fourier transformed and the 
amplitude spectra searched for significant features. To increase 
sensitivity to narrow duty-cycle pulses, individual spectra summing the 
first 2, 4, 8 and 16 harmonics were also searched. The resulting 
list of candidate signals above a S/N threshold of 6 were then 
folded in the time domain for visual inspection. No convincing 
pulsar-like signals were found.

Based on the known system parameters (Camilo et al.~2002) we estimate the
sensitivity\footnote{We note that the radio limit which we have calculated 
are valid only at beam center. At the 15'-FWHM point, the sensitivity is 
estimated to be a factor of two lower.} of our observations to 
be $S_{\rm min}= 0.2\,\delta^{1/2}$ mJy 
at periods $\gtrsim 10$ ms and DMs consistent with the 1.5 kpc distance. 
With a typical duty cycle $\delta=0.04$, this yields $S_{\rm min} \sim 
40\,\mu$Jy. For a distance of 1.5 kpc the 820-MHz luminosity limit 
$L_{820}$ is $0.09\,\mbox{mJy kpc}^2$. For an assumed radio spectral 
index of $-1.0$ (see e.g.~Lorimer et al.~1995) the corresponding luminosity 
at 1400 MHz is $L_{1400} \sim 0.05\,\mbox{mJy kpc}^2$ which is lower than 
most of the recent deep radio search limits (see e.g.~Camilo 2003).
If there is a pulsar operating in \EG, our sensitivity limit suggests that the
pulsar either does not produce significant amounts of radio emission or that its
geometry is such that the radio beam does not intersect with the line of sight.

\section{Discussion and Summary}

We have searched a portion of the $\gamma$-Cygni field for possible X-ray
counterparts to the intriguing gamma-ray source \EG\, (\CG) using {\em Chandra} 
and  ROSAT. We have shown that a previous candidate, \RX, is almost certainly 
not the gamma-ray source but identified with a normal star. This conclusion 
is based on the refined position of the X-ray source, its spectrum and 
coincidence with both optical and infrared sources and the inferred X-ray 
luminosity. Further, we have found a total of 38 X-ray sources in the 
{\em Chandra} S2-, S3- and S4-fields which covers only part of the much larger 
error box containing the location of the EGRET source. A re-analysis of
archival ROSAT HRI data revealed three more X-ray sources within the EGRET
error box which are not detected in the {\em Chandra} observations. Two of 
these sources are surely variable whereas the third source was found in a 
region not covered by the {\em Chandra} observation.  We found  that some 
of the {\em Chandra} sources have counterparts that may be main-sequence 
stars based on their identification with optical objects and 2MASS sources 
of normal colors. Of course the X-ray emission may not be due to the main 
sequence star, but can arise from an accreting compact companion.  None 
of the X-ray sources appear to be radio pulsars, down to a limiting 
sensitivity of $L_{820} = 0.09\, \mbox{mJy kpc}^2$ for an assumed pulse 
duty cycle of 4\%. This limit also applies to the entire region associated 
with the 99\%-confidence position contours of \EG. 

Determining an upper limit for a putative X-ray point source located
within the 68\% confidence contour of \EG\, using archival ROSAT PSPC 
data we found a $2\sigma$ luminosity upper limit of $L_x(\mbox{0.1-2.4 keV}) 
< 4.8 \times 10^{31}\,\mbox{erg}\,\mbox{s}^{-1}$ which is four times smaller 
than the ROSAT PSPC-deduced luminosity observed from the Vela pulsar 
($L_x = 1.77 \times 10^{32}\, \mbox{erg}\,\mbox{s}^{-1}$, e.g.~Table 3 in 
Becker \& Aschenbach 2002) but about four times higher than the total  ROSAT 
observed X-ray luminosity from Geminga ($L_x = 1.26 \times 10^{31}\, 
\mbox{erg}\,\mbox{s}^{-1}$). We therefore consider it as a valid option
that the counterpart of \EG\, is a neutron star with an X-ray luminosity similar
to that observed from Vela-like to middle-aged pulsars. An object with 
such luminosity would not have been detected in the X-ray data from ROSAT 
which cover that region of the sky. Adopting the SNR age of 5400 yrs and assuming
a spin-down to X-ray energy conversion factor of $10^{-4}$ (Becker \& Tr\"umper
1997) we are able to constrain the spin-parameters of such a putative neutron
star to be $P \gtrsim 160\, (d/1.5\, \mbox{kpc})^{-1}\mbox{ms}$, $\dot{P} 
\gtrsim 5\times 10^{-13}\, (d/1.5 \mbox{kpc})^{-1}\,\mbox{s s}^{-1}$ and
$B_\perp \gtrsim 9\times 10^{12}\, (d/1.5\,\mbox{kpc})^{-1}$ G, which are 
consistent with the properties of known Vela- to middle-aged pulsars
(e.g.~Gonzalez \& Safi-Harb 2003, Becker \& Pavlov 2001), given
the uncertainty of this approach. The ratio of the $\gamma$-ray to soft 
X-ray flux deduced from our upper limit, $f_\gamma/f_x > 2400$, is 
consistent with this conclusion.

In order to obtain a full census of the X-ray population in the error
box of 3EG J2020+4017 further observations with {\em Chandra} are required. As
the discovery of Geminga has taught us, deep follow-up optical
observations of new X-ray sources can also lead to the identification of the 
nature of a high-energy source. Finally, the
measurements we expect from the {\em GLAST} mission (launch 2007) should
provide a much improved signal-to-noise ratio and a source location
better than $0\farcm5$ for this gamma-ray source. This will open the
possibility to directly search for pulsar periodicities in the
gamma-ray data. In case no pulsar is found, the restricted number of
{\em Chandra} sources in the {\em GLAST} error box will then be prime candidates for
even deeper searches for counterparts.

\acknowledgments
Those of us at the Marshall Space Flight Center acknowledge support from the
{\em Chandra} Project. MCW acknowledges with gratitude conversations with 
Marshall Joy and Roc Cutri that clarified some of the mysteries of the infrared 
portion of the spectrum. ZA was supported by NASA grant NRA-99-01-LTSA-070. DRL 
is a University Research Fellow funded by the Royal Society. FC is supported 
in part by NSF grant AST-02-05853.
This publication makes use of data products from the Two Micron All Sky Survey,
which is a joint project of the University of Massachusetts and the Infrared
Processing and Analysis Center/California Institute of Technology, funded by
the National Aeronautics and Space Administration and the National Science
Foundation. In addition, this research has made use of data obtained from 
the High Energy Astrophysics Science Archive Research Center (HEASARC),
provided by NASA's Goddard Space Flight Center.

\begin{deluxetable}{lrrrrrrrrc}
\tabletypesize{\scriptsize}
   \tablewidth{0pc}
   \tablecaption{{\em Chandra} X-ray sources in the $\gamma$-Cygni field.\label{GamCyg_source_table}}
    \tablehead{}
   \startdata

   \multicolumn{1}{l}{NAME}
 & \multicolumn{1}{c}{R.A.}       & \multicolumn{1}{c}{Dec.}
 & \multicolumn{1}{c}{$r_1$$^a$}  & \multicolumn{1}{c}{$N^b$}
 & \multicolumn{1}{c}{$S/N$$^c$}  & \multicolumn{1}{c}{$r_2$$^d$}
 & \multicolumn{1}{c}{USNO$^e$}   & \multicolumn{1}{c}{2MASS$^e$}
 & \multicolumn{1}{c}{ ROSAT$^e$}    \\
 &  \multicolumn{1}{c}{(J2000)}   & \multicolumn{1}{c}{(J2000)}
 & ($\arcsec$) & & &($\arcsec$) & ($\arcsec$) & ($\arcsec$) & ($\arcsec$)\\\hline
  S401 & 304.86758 & 40.436279 & 13.6 & 49  & 5.7 & 1.25 & 0.61\&1.55 &      & \\
  S402 & 304.88129 & 40.373795 & 13.8 & 25  & 3.9 & 1.72 &      & 1.65 & \\
  S403 & 304.92621 & 40.362309 & 10.1 & 69  & 6.7 & 0.87 & 1.05 & 1.13 &  \\
  S301 & 304.97339 & 40.455490 & 4.54 & 13  & 3.0 & 0.88 & 0.70 & 0.63 & \\
  S302 & 304.97769 & 40.468494 & 4.61 & 36  & 5.2 & 0.65 &      &      & \\
  S303 & 304.99615 & 40.398434 & 3.75 & 14  & 3.1 & 0.76 & 0.85 & 1.29 & \\
  S304 & 305.00381 & 40.365112 & 5.03 & 14  & 3.0 & 0.94 & 1.15 & 1.41 & \\
  S305 & 305.00769 & 40.434021 & 2.63 & 176 & 11.2 & 0.47 &      & 0.56 & \\
  S306 & 305.01837 & 40.457535 & 2.47 & 23  & 4.3 & 0.55 &      &      &\\
  S307 & 305.02713 & 40.412449 & 2.19 & 25  & 4.5 & 0.52 & 0.37 & 0.39 & \\
  S308 & 305.02737 & 40.374905 & 3.60 & 28  & 4.8 & 0.61 & 0.25 & 0.36 & \\
  S309 & 305.05789 & 40.394272 & 2.04 & 15  & 3.3 & 0.55 &      & 0.17 & \\
  S310 & 305.06354 & 40.486912 & 2.46 & 12  & 3.1 & 0.63 & 0.95 & 1.05 & \\
  S311 & 305.07104 & 40.446281 & 1.21 & 44  & 5.8 & 0.47 & 0.30 & 0.25 & \\
  S312 & 305.07147 & 40.437424 & 1.14 & 253 & 13.7 & 0.46 & 0.27 & 0.30 & 5.9$^f$ \\
  S313 & 305.08884 & 40.448517 & 1.21 & 20  & 3.7 & 0.48 &      &      & \\
  S314 & 305.09262 & 40.490631 & 2.57 & 22  & 4.0 & 0.56 &      & 0.13 & \\
  S315 & 305.10596 & 40.484375 & 2.37 & 12  & 3.0 & 0.61 & 0.65 & 0.41 & \\
  S316 & 305.11560 & 40.440331 & 1.40 & 11  & 3.1 & 0.52 &      &      & \\
  S201 & 305.14136 & 40.473434 & 2.68 & 10  & 2.9 & 0.68 & 0.34 & 0.62 & \\
  S202 & 305.15140 & 40.494373 & 3.97 & 17  & 3.5 & 0.73 & 0.80 & 0.64 & \\
  S203 & 305.15570 & 40.495060 & 4.18 & 11  & 2.8 & 0.88 &      &      & \\
  S204 & 305.17081 & 40.451115 & 3.26 & 85  & 7.9 & 0.50 & 0.22 & 0.35 & 1.2$^g$\\
  S205 & 305.17349 & 40.380219 & 4.65 & 29  & 4.7 & 0.69 & 0.73 &      & \\
  S206 & 305.18216 & 40.450073 & 3.82 & 213 & 12.4& 0.48 &      & 0.31 & \\
  S207 & 305.18600 & 40.430450 & 3.93 & 10  & 2.9 & 0.87 & 0.78 & 1.03 & \\
  S208 & 305.21335 & 40.403984 & 6.05 & 16  & 3.4 & 1.02 &      & 0.84 & \\
  S209 & 305.21414 & 40.509041 & 8.20 & 28  & 4.2 & 1.04 & 1.73 & 1.66 & \\
  S210 & 305.21881 & 40.473904 & 6.70 & 24  & 3.8 & 0.95 & 0.55 & 0.55 & \\
  S211 & 305.21906 & 40.408849 & 6.35 & 66  & 7.0 & 0.65 & 0.43 & 0.62 & \\
  S212 & 305.22256 & 40.507294 & 8.70 & 47  & 5.9 & 0.89 & 1.54 & 1.61 & \\
  S213 & 305.23041 & 40.474827 & 7.60 & 33  & 4.2 & 0.95 & 0.91 & 0.63 & \\
  S214 & 305.24078 & 40.474091 & 8.44 & 101 & 8.2 & 0.68 & 0.60 & 0.40 & 2.9$^h$\\
  S215 & 305.24927 & 40.378075 & 9.95 & 17  & 3.1 & 1.52 & 0.91 & 0.95 & \\
  S216 & 305.25110 & 40.485874 & 9.80 & 33  & 4.3 & 1.12 & 1.63 & 1.10 & \\
  S217 & 305.27808 & 40.430069 & 11.2 & 24  & 3.8 & 1.47 & 0.97 & 0.97 & \\
  S218 & 305.29422 & 40.513210 & 15.8 & 35  & 4.0 & 1.66 &      & 0.57 & \\
  S219 & 305.29767 & 40.468163 & 13.9 & 201 &11.5 & 0.74 & 0.38 & 0.70 & 2.4$^i$\\
%
\enddata
 \tablecomments{\\
$^a$  Extraction radius. \\
$^b$  Approximate number of source counts.\\
$^c$  Detection signal-to-noise ratio.\\
$^d$  X-ray position uncertainty ($1\sigma$ radius).\\
$^e$  Radial separation between X-ray position and cataloged position of counterpart.\\
$^f$ 1RXH J2020.2+4026; the positional uncertainty is 6 arcsec.\\ 
$^g$ 1RXH J202040.9+402704; the positional uncertainty is larger than the separation (2-3 arcsec)\\
$^h$ 1RXH J202057.8+402829; the positional uncertainty is 1 arcsec.\\
$^i$ 1RXH J202111.4+402807; the positional uncertainty is 2 arcsec.\\
}
 \end{deluxetable}

\begin{deluxetable}{lrrrrrrc}
\tabletypesize{\scriptsize}
   \tablewidth{0pc}
   \tablecaption{Candidate counterparts to the X-ray sources in the $\gamma$-Cygni field.\label{t:counterparts}}
    \tablehead{}
   \startdata

   \multicolumn{1}{l}{NAME}
 & \multicolumn{1}{c}{R.A.}       & \multicolumn{1}{c}{Dec.}
 & \multicolumn{1}{c}{$N_{r99}$$^a$}  & \multicolumn{1}{c}{R.A.}
 & \multicolumn{1}{c}{Dec.}  & \multicolumn{1}{c}{$P_{r99}$$^a$}
 & \multicolumn{1}{c}{$\delta$$^b$}  \\
 &  \multicolumn{1}{c}{(J2000)}   & \multicolumn{1}{c}{(J2000)}
 &  & \multicolumn{1}{c}{(J2000)}& \multicolumn{1}{c}{(J2000)}& & ($\arcsec$) \\
 &  \multicolumn{1}{c}{USNO}   & \multicolumn{1}{c}{USNO}
 &                             & \multicolumn{1}{c} {2MASS} 
 & \multicolumn{1}{c} {2MASS}  & &  \\\hline
  S401 & 304.867803 & 40.436275 & 0.041 &            &           &       &  \\
  S401 & 304.868070 & 40.436495 & 0.041 &            &           &       &  \\
  S402 &            &           &       & 304.881105 & 40.373360 & 0.117 &  \\
  S403 & 304.926575 & 40.362225 & 0.020 & 304.926589 & 40.362183 & 0.030 & 0.16 \\
  S301 & 304.973164 & 40.455400 & 0.021 & 304.973295 & 40.455330 & 0.031 & 0.44 \\
  S303 & 304.996045 & 40.398212 & 0.015 & 304.995913 & 40.398125 & 0.023 & 0.48  \\
  S304 & 305.003723 & 40.364800 & 0.023 & 305.003740 & 40.364723 & 0.035 & 0.28 \\
  S305 &            &           &       & 305.007841 & 40.434124 & 0.009 &  \\
  S307 & 305.027000 & 40.412425 & 0.007 & 305.027043 & 40.412365 & 0.011 & 0.25 \\
  S308 & 305.027320 & 40.374964 & 0.010 & 305.027281 & 40.374832 & 0.015 & 0.49 \\
  S309 &            &           &       & 305.057841 & 40.394245 & 0.012 & \\
  S310 & 305.063206 & 40.486845 & 0.010 & 305.063218 & 40.486755 & 0.015 & 0.33 \\
  S311 & 305.070998 & 40.446359 & 0.006 & 305.071003 & 40.446217 & 0.009 & 0.51 \\
  S312 & 305.071387 & 40.437464 & 0.005 & 305.071389 & 40.437366 & 0.008 & 0.35 \\
  S314 &            &           &       & 305.092591 & 40.490601 & 0.012 & \\
  S315 & 305.105898 & 40.484550 & 0.010 & 305.105912 & 40.484482 & 0.015 & 0.25 \\
  S201 & 305.141242 & 40.473467 & 0.012 & 305.141188 & 40.473324 & 0.018 & 0.54     \\
  S202 & 305.151131 & 40.494459 & 0.014 & 305.151193 & 40.494289 & 0.021 & 0.64 \\
  S204 & 305.170775 & 40.451170 & 0.007 & 305.170747 & 40.451031 & 0.010 & 0.51 \\
  S205 & 305.173353 & 40.380392 & 0.013 &            &           &       & \\
  S206 &            &           &       & 305.182074 & 40.450016 & 0.009 & \\
  S207 & 305.186253 & 40.430348 & 0.020 & 305.186255 & 40.430241 & 0.030 & 0.39 \\
  S208   &            &           &     & 305.213189 & 40.403786 & 0.041 & \\
  S209 & 305.214453 & 40.508625 & 0.028 & 305.213771 & 40.508675 & 0.042 & 1.88 \\
  S210 & 305.218609 & 40.473920 & 0.024 & 305.218624 & 40.473850 & 0.035 & 0.26 \\
  S211 & 305.219192 & 40.408425 & 0.011 & 305.219198 & 40.408352 & 0.017 & 0.26 \\
  S212 & 305.223092 & 40.507159 & 0.021 & 305.223083 & 40.507088 & 0.031 & 0.26 \\
  S213 & 305.230259 & 40.475053 & 0.024 & 305.230230 & 40.474934 & 0.036 & 0.44 \\
  S214 & 305.240995 & 40.474125 & 0.012 & 305.240870 & 40.474003 & 0.018 & 0.56 \\
  S215 & 305.249534 & 40.378228 & 0.061 & 305.249506 & 40.378269 & 0.090 & 0.17 \\
  S216 & 305.250842 & 40.486821 & 0.033 & 305.250918 & 40.486145 & 0.049 & 0.53 \\
  S217 & 305.278420 & 40.429992 & 0.057 & 305.278386 & 40.429932 & 0.085 & 0.24 \\
  S218 &            &           &       & 305.294173 & 40.513363 & 0.109 & \\
  S219 & 305.297639 & 40.468059 & 0.015 & 305.297601 & 40.467976 & 0.022 & 0.32 \\ 
%
\enddata
 \tablecomments{\\
$^a$  The average number of accidential coincidences expected in the region searched. \\
$^b$  Angular separation between the USNO and 2MASS candidate counterpart. \\
}
 \end{deluxetable}

\clearpage

\begin{deluxetable}{ccccccccccccccc}
\tabletypesize{\scriptsize}
  \tablewidth{0pc}
  \tablecaption{2MASS counterparts: magnitudes and colors.\label{t:2MASS_colors}}
  \tablehead{}
  \startdata
Source & J & $\sigma$ & H & $\sigma$ & K$_S$ & $\sigma$ & J$-$H & $\sigma$&  H$-$K$_S$ & $\sigma$&  J$-$K$_S$ & $\sigma$  \\ \hline\\[-1.5ex]
S201 & 14.821 & 0.038 & 14.136 & 0.053 & 13.706 & 0.061 & 0.685   & 0.065 & 0.430  & 0.081 & 1.115  & 0.072 \\
S202 & 14.458 & 0.038 & 13.702 & 0.037 & 13.485 & 0.05  & 0.756   & 0.053 & 0.217  & 0.062 & 0.973  & 0.063 \\
S204 & 13.243 & 0.029 & 12.566 & 0.032 & 12.359 & 0.036 & 0.677   & 0.043 & 0.207  & 0.048 & 0.884  & 0.046 \\
S206 & 14.516 & 0.051 & 13.921 & 0.051 & 13.796 & 0.069 & 0.595   & 0.072 & 0.125  & 0.086 & 0.72   & 0.086 \\
S207 & 15.235 & 0.052 & 14.678 & 0.070 & 14.58  & 0.12  & 0.557   & 0.087 & 0.098  & 0.139 & 0.655  & 0.131 \\
S208 & 17.597$^a$ &       & 15.246 & 0.133 & 14.614 & 0.12  & $>2.4$  &       & 0.632  & 0.179 & $>3.0$ &       \\
S209 & 14.281 & 0.052 & 13.232 & 0.068 & 12.725 & 0.063 & 1.049   & 0.086 & 0.507  & 0.093 & 1.556  & 0.082 \\
S210 & 14.648 & 0.038 & 13.685 & 0.039 & 13.249 & 0.043 & 0.963   & 0.054 & 0.436  & 0.058 & 1.399  & 0.057 \\
S211 & 13.503 & 0.026 & 12.695 & 0.021 & 12.338 & 0.03  & 0.808   & 0.033 & 0.357  & 0.037 & 1.165  & 0.040 \\
S212 & 11.468 & 0.022 & 11.075 & 0.020 & 10.9   & 0.023 & 0.393   & 0.030 & 0.175  & 0.030 & 0.568  & 0.032 \\
S213 & 14.99  & 0.045 & 14.151 & 0.052 & 13.802 & 0.063 & 0.839   & 0.069 & 0.349  & 0.082 & 1.188  & 0.077 \\
S214 & 12.674$^a$ &   & 11.851 & 0.035 & 11.606 & 0.039 & $>0.8$  &       & 0.245  & 0.052 & $>1.1$ &       \\
S215 & 14.361 & 0.036 & 13.29  & 0.036 & 11.96$^a$  &       & 1.071   & 0.051 & $< 1.3$   &    & $< 2.4$  &  \\
S216 & 13.736 & 0.028 & 12.929 & 0.031 & 12.615 & 0.034 & 0.807   & 0.042 & 0.314  & 0.046 & 1.121  & 0.044 \\
S217 & 11.733 & 0.021 & 11.497 & 0.018 & 11.463 & 0.018 & 0.236   & 0.028 & 0.034  & 0.025 & 0.27   & 0.028 \\
S218 & 14.9$^a$  &       & 15.591 & 0.136 & 13.743$^a$ &       & $>-0.7$ &       & $<1.9$ &       & N/A    &       \\
S219 & 11.927 & 0.021 & 11.252 & 0.018 & 11.155 & 0.017 & 0.675   & 0.028 & 0.097  & 0.025 & 0.772  & 0.027 \\
S301 & 14.915 & 0.039 & 13.742 & 0.037 & 13.337 & 0.042 & 1.173   & 0.054 & 0.405  & 0.056 & 1.578  & 0.057 \\
S303 & 15.776 & 0.069 & 14.678 & 0.064 & 14.283 & 0.091 & 1.098   & 0.094 & 0.395  & 0.111 & 1.493  & 0.114 \\
S304 & 15.764 & 0.068 & 14.914 & 0.076 & 14.479 & 0.103 & 1.750   & 0.102 & -0.465 & 0.128 & 1.285  & 0.123 \\
S305 & 11.185 & 0.023 & 10.783 & 0.018 & 10.673 & 0.016 & 0.402   & 0.029 & 0.11   & 0.024 & 0.512  & 0.028 \\
S307 & 14.559 & 0.033 & 13.225 & 0.026 & 12.636 & 0.026 & 1.334   & 0.042 & 0.589  & 0.037 & 1.923  & 0.042 \\
S308 & 14.29  & 0.039 & 13.375 & 0.063 & 13.007 & 0.036 & 0.915   & 0.074 & 0.368  & 0.073 & 1.283  & 0.053 \\
S309 & 17.863$^a$ &       & 15.472 & 0.118 & 14.683 & 0.125 & $>2.4$  & 0.118 & 0.789  & 0.172 & $>3.2$ &       \\
S310 & 15.37  & 0.052 & 14.331 & 0.049 & 14.127 & 0.076 & 1.039   & 0.071 & 0.204  & 0.090 & 1.243  & 0.092 \\
S311 & 15.181 & 0.054 & 14.246 & 0.054 & 13.957 & 0.067 & 0.935   & 0.076 & 0.289  & 0.086 & 1.224  & 0.086 \\
S312 & 12.373 & 0.022 & 11.83  & 0.021 & 11.686 & 0.017 & 0.543   & 0.030 & 0.144  & 0.027 & 0.687  & 0.028 \\
S314 & 16.492 & 0.135 & 15.4   & 0.107 & 14.906 & 0.158 & 1.092   & 0.172 & 0.494  & 0.191 & 1.586  & 0.208 \\
S315 & 15.365 & 0.052 & 14.738 & 0.062 & 14.38  & 0.101 & 0.627   & 0.081 & 0.358  & 0.119 & 0.985  & 0.114 \\
S402 & 14.456 & 0.038 & 13.702 & 0.035 & 13.477 & 0.045 & 0.754   & 0.052 & 0.225  & 0.057 & 0.979  & 0.059 \\
S403 & 13.262 & 0.023 & 12.749 & 0.024 & 12.634 & 0.028 & 0.513   & 0.033 & 0.115  & 0.037 & 0.628  & 0.036 \\
 \enddata
 \tablecomments{\\
$^a$  Lower limit. \\
}
 \end{deluxetable}

\clearpage

\begin{deluxetable}{cccccc}
  \tablewidth{0pc}
  \tablecaption{Spectral fits to the brightest sources.\label{t:spectra}}
  \tablehead{}
  \startdata
Source &  Model$^a$ & \chisq & $\nu$ & $N_H/10^{22}$ & $\Gamma$ or kT (keV)  \\ 
{}     &    {}      &   {}   &  {}   & $\mbox{cm}^{-2}$ &      {}      \\  \hline\\[-1.5ex]
S312 & PL       & 16.6 & 15 & 1.07 (0.80$ - $1.45) & 8.2  (6.8$ - $10.0)  \\
S312 & MEKAL    & 22.3 & 15 & 0.00 (0.00$ - $0.05) & 0.77 (0.68$ - $0.83) \\
S312 & BB       & 14.6 & 15 & 0.37 (0.22$ - $0.61) & 0.14 (0.12$ - $0.15) \\[1ex]

S206 & PL       & 18.7 & 16 & 0.01 (0.00$ - $0.15) & 1.92 (1.67$ - $2.36) \\
S206 & MEKAL    & 43.4 & 16 & 1.2                & 1.0                \\
S206 & BB       & 32.8 & 16 & 0.0                & 0.5                \\[1ex]

S219 & PL       & 29.8 & 16 & 0.5                & 5.4                \\
S219 & MEKAL    & 49.2 & 16 & 0.0                & 1.0                \\
S219 & BB       & 33.9 & 16 & 0.0                & 0.2               \\[1ex]

S305 & PL       & 18.2 & 12 & 0.48 (0.16$ - $0.82) & 2.98 (2.51$ -$ 3.89) \\
S305 & MEKAL    & 17.8 & 12 & 1.28 (1.07$ - $1.51) & 0.70 (0.59$- $ 0.82)   \\
S305 & BB       & 21.7 & 12 & 0.0  (0.0$ -  $0.20) & 0.39 (0.32$- $ 0.44)   \\[1ex]

S214 & PL       & 12.3 & 6  & 0.2                & 3.2                \\
S214 & MEKAL    &  9.7 & 6  & 0.9                & 0.9                \\
S214 & BB       & 14.6 & 6  & 0.0                & 0.3                \\[1ex]

S204 & PL       & 7.75 & 4  & 0.0                & 2.6                \\
S204 & MEKAL    & 11.6 & 4  & 1.0                & 1.0                \\
S204 & BB       & 12.0 & 4  & 0.0                & 0.4                \\
 \enddata
\tablecomments{$^a$BB = blackbody; PL=power law. Uncertainties for 
statistically unacceptable fits are not quoted. For more details see 
\S \ref{S312} and \S\ref{other_bright_sources}.}. 
 \end{deluxetable}


\begin{deluxetable}{ccccc}
  \tablewidth{0pc}
  \tablecaption{Pointing centers of the ROSAT HRI observations which partly cover the error box of \EG.\label{hri}}
  \startdata
    Seq.    & Start date & Exposure &      R.A. (J2000)   & Dec. (J2000)\\ 
    number  &   (YMD)    &  (sec)   &       (HMS)      &  (DMS)  \\ \hline\\[-1.5ex]
    202033h &  1994 11 15  & 10536    &  20 20 28.08 & 41 21 36 \\ 
    202534h &  1997 06 06  & 10370    &  20 22 14.04 & 40 15 36 \\
    400899h &  1996 11 02  & 36552    &  20 21 04.08 & 40 26 24 \\
    500339h &  1994 11 14  & 18761    &  20 19 48.00 & 40 03 00 \\
 \enddata
 \end{deluxetable}

\begin{deluxetable}{cccccc}
  \tablewidth{0pc}
  \tablecaption{ ROSAT HRI sources in G78.2+2.1.\label{hri_sources}}
  \startdata
 Data    &    Source  & R.A. (J2000)   & Dec. (J2000) & HRI rate &   {\em Chandra} \\
  {}     &   {}     &      {}          &     {}      &  cts/s $\times 10^{-4}$ & {}   \\ \hline\\[-1.5ex]
 400899h & RX J202137.6+402959  &  305.407035  & 40.499979    &  $8.0 \pm 2.0$  & ---   \\
 400899h & RX J202057.8+402829  &  305.240996  & 40.474830    &  $7.9 \pm 1.6$  & S214  \\
 400899h & RX J202111.3+402806  &  305.297479  & 40.468571    &  $16.5 \pm 2.3$ & S219  \\
 400899h & RX J202040.8+402704  &  305.170322  & 40.451285    &  $6.6 \pm 1.6$  & S204  \\
 400899h & RX J202130.5+402649  &  305.377281  & 40.447124    &  $6.1 \pm 1.6$  & ---   \\
 400899h & RX J202016.8+402614  &  305.070074  & 40.437480    &  $13.1 \pm 2.4$ & S312  \\

 202534h & RX J202240.0+401900  &  305.666926  & 40.316753    &  $16.7 \pm 4.3$ & ---   \\
 202534h & RX J202150.5+401837  &  305.460420  & 40.310447    &  $20.7 \pm 5.0$ & ---   \\

 500339h & RX J201950.8+395752  &  304.961886  & 39.964551    &  $33.0 \pm 4.5$ & ---   \\
 \enddata
 \end{deluxetable}


\begin{deluxetable}{ccc}
  \tablewidth{0pc}
  \tablecaption{Pointing centers of the four radio observations.\label{radio_pointings_tabel}}
  \startdata
  R.A. (J2000) &  Dec. (J2000) & Dwell \\ 
   (HMS)       &     (DMS)     &  (hour) \\ \hline\\[-1.5ex]
20 20 18.07    & 40 24 35.1   &  4  \\  
20 21 21.53    & 40 23 21.7   &  4  \\
20 21 25.06    & 40 14 47.1   &  4  \\
20 20 33.07    & 40 13 15.7   &  3  \\
 \enddata
 \end{deluxetable}

\begin{thebibliography}{}

\bibitem[]{} Backer, D.C., Dexter, M.R., Zepka, A., Ng, D., Werthimer, D.J., 
Ray, P.S., Foster, R.S., 1997, PASP, 109, 61

\bibitem[]{} Becker, W., Aschenbach, B., 2002, in {Proceedings of the
WE-Heraeus Seminar on Neutron Stars, Pulsars and Supernova remnants},
Eds.~W.Becker, H.Lesch \& J/Tr\"umper, MPE-Report 278, 64, (available from astro-ph/0208466)

\bibitem[]{} Becker, W., Pavlov, G.G., 2001, in {\em The Century of Space  
Science}, eds. J.Bleeker, J.Geiss \& M.Huber, Kluwer Academic Publishers,  
p721 (available from astro-ph/0208356).

\bibitem[]{} Becker, W., Tr\"umper, J., 1997, A\&A, 326, 682

\bibitem[]{} Bertsch, D.L., Hartman, R.C., Hunter, S.D. et al., 2000, in {\em
Proc.~5th
Compton Symposium}, AIP-CP 510, p504

\bibitem[]{} Brazier, K.T.S., Kanbach, G., Carraminana, A., et al, 1996, MNRAS,
281, 1033

\bibitem[]{} Camilo, F., Bell, J.F., Manchester, R.N., Lyne, A.G., Possenti, A.,
Kramer, M., Kaspi, V.M., Stairs, I.H., D'Amico, N., Hoobs, G., Gotthelf, E.V.,
Gaensler, B.M., 2001, ApJ, 557, L51


\bibitem[]{} Camilo, F., Stairs, I.H., Lorimer, D.R., Backer, D.C., Ransom, S.M., 
Klein, B., Wielebinski, R., Kramer, M., McLaughlin, M.A., Arzoumanian, Z., M\"uller, P., 2002, ApJ,  571, L41

\bibitem[]{} Camilo, F., 2003, in {\em  Radio Pulsars}, Eds M.Bailes, D.J.Nice, and S.E.Thorsett, 
Astronomical Society of the Pacific, San Francisco

\bibitem[]{} Carraminana, A.,  Chavushyan, V., Zharikov, S., et al., 2000, in
{\em Proc. 5th Compton Symposium}, AIP-CP 510, p49

\bibitem[]{} Cordes, J.M., Lazio, T.J.W., 2002, astro-ph/0207156

\bibitem[]{} D'Amico, N.,  Kaspi, V.M., Manchester, R.N.,  et al. 2001 ApJ 552, L45

\bibitem[]{} Dickey, J. M., Lockman, F. J., 1990, Ann.~Rev.~Astron.~Astrophys., 28, 215

\bibitem[]{} Fleming, T.A., Molendt, S., Maccacaro, T., Woltjer, A., 1995, ApJS, 99, 701

\bibitem[]{} Green, D.A., 1989, A\&AS, 78, 277

\bibitem[]{} Gil, J.A., Khechinashvili, D.G., Melikidze, G.I., 1998, MNRAS, 298, 1207

\bibitem[]{} Gonzalez, M., Safi-Harb, S., 2003, ApJ, 591, 143

\bibitem[]{} Halpern, J.P., Gotthelf, E.V., Mirabal, N., Camilo, F., 2002, ApJ, 573, L41

\bibitem[]{} Halpern, J.P., Eracleous, M., Mukherjee, R.,  et al. 2001a, ApJ
551, 1016

\bibitem[]{} Halpern, J.P.,  Camilo, F., Gotthelf, E. V., et al. 2001b, ApJ 552,
L125

\bibitem[]{} Hartman, R.C.,  Bertsch, D.L., Bloom, S.D., et al., 1999, ApJS, 123, 79

\bibitem[]{} Hessels, J.W.T., Roberts, M.S.E., Ransom, S.M., Kaspi, V.M., Romani, R.W., 
Ng, C.Y., Freier, P.C.C., Gaensler, B.M., 2004, astro-ph/0403632

\bibitem[]{} Higgs, L.A., Landecker, T.L., and Roger, R.S., 1977, AJ, 82, 718

\bibitem[]{} Kanbach, G., Bertsch, D.L., Dingus, B.L., et al., 1996, in
{\em Proc.~3rd Compton Symposium}, A\&AS, 120, 461

\bibitem[]{} Kanbach, G., 2002, in Proc.~of the 270. WE-Heraeus Seminar on Neutron
Stars, Pulsars and Supernova Remnants, eds.~W.Becker, H.Lesch and J-Tr\"umper. MPE-Report 278,
p91 (astro-ph/0209021)

\bibitem[]{} Kramer, M.,  Bell, J.F., Manchester, R.N.  et al., 2003, MNRAS 342, 1299

\bibitem[]{} Kuzmin, A. D.; Losovkii, B. Y., Pisma Astron. Zh., 1997, 23, 323

\bibitem[]{} Landecker, T.L., Roger. R.S., and Higgs, L.A., 1980, A\&AS, 39, 133

\bibitem[]{} Lorimer, D.R., Kramer, M., M\"uller, P., Wex, N., Jessner, A., 
Lange, C., Wielebinski, R., 2000, A\&A,  358, 169

\bibitem[]{} Lorimer, D.R., Yates, J.A., Lyne, A.G., Gould, D.M., 1995, MNRAS, 273, 411L

\bibitem[]{} Maeda, Y., Koyama, K., Yokogawa, J., Skinner, S., 1999, ApJ 510, 967

\bibitem[]{} Mayer-Hasselwander, H.A., Simpson, G., 1990, in {\em the EGRET
Science Symposium},
eds C.Fichtel, S.Huntre, P.Sreekumar, F.Stecker, NASA, VP-3071, p153

\bibitem[]{} Merck, M., Bertsch, D.L., Dingus, B.L.,  et al.,~1996,  in {\em
Proc. 3rd Compton Symposium}, A\&AS, 120, 465

\bibitem[]{} Mirabal, N., Halpern, J.P., 2001, ApJ 547, L137

\bibitem[]{} Monet, D.G., Levine, S.E., Canzian, B., et al., 2003, AJ, 125,984

\bibitem[]{} Nice, D.J., Sayer, R.W.,1997, ApJ, 476, 261

\bibitem[]{} Reimer, O., Brazier, K. T. S., Carraminana, A., et al., 2001, MNRAS, 324, 772

\bibitem[]{} Reimer, O., Bertsch, D.L., 2001, in Proc.~27th ICRC (Hamburg), Ed. M.Simon, 
E.Lorenz, M.Pohl, Vol.6, p2546


\bibitem[]{} Roberts, M.S.E., Hessels, J.W.T., Ransom, S.M., Kaspi, V.M., Freire, P.C.C., 
Crawford, F., Lorimer, D.R., 2002, ApJ, 577, L19

\bibitem[]{} Roberts, M., Ransom, S., Hessel, J., et al, 2004, in {\em Young
Neutron Stars and their environments}, IAU Symposium, Vol 218, eds. F.Camilo and
B.Gaensler

\bibitem[]{} Romero, G.E., Benaglia, P., Torres, D. F., et al., 2000, in
{\em Proc. 5th Compton Symposium}, AIP-CP 510, p509

\bibitem[]{} Stocke, J.T., Morris, S.L., Gioia, et al.,1991, ApJS, 76, 813

\bibitem[]{} Sturner, S.J., Dermer, C.D., 1995, A\&A, 293, L17

\bibitem[]{} Swanenburg, B.N., Bennett, K., Bignami, G. F., et al.,1981, ApJ,
243, L69

\bibitem[]{} Swartz, D. A., Ghosh, K.K., McCollough, M.L., Pannuti, T.G.,
Tennant, A.F. \& Wu, K. 2003, ApJS, 144, 213

\bibitem[]{} Thompson, D.J., Fichtel, C. E., Kniffen, D. A., et al, 1975, ApJ,
200, L79

\bibitem[]{} Torres D.F., Butt, Y.M., Camilo, F., 2001, ApJ 560, L155

\bibitem[]{} Uchiyama Y., Takahashi, T., Aharonian, F.A., Mattox, J.R., 2002, ApJ, 571, 866

\bibitem[]{} Wallace P.M., Halpern, J.P., Magalhaes, A.M., et al., 2002 ApJ,
569, 36

\bibitem[]{} Wendker, H.J., Higgs, L.A., and Landecker, T.L., 1991, A\&A, 241, 551

\bibitem[]{} Wilms,J., Allen,A. \& McCray,R. 2000, ApJ, 542, 914

\bibitem[]{} Yadigaroglu, I.-A., and Romani, R.W., 1995, ApJ, 449, 211

\end{thebibliography}
\end{document}